\documentclass[10pt]{iopart}
\usepackage{iopams}
\usepackage{theorem}
\usepackage[dvips]{graphicx,epsfig}
\usepackage{amssymb,bbm}
\usepackage{euscript}

\newcommand{\B}{\mathfrak{B}}

\newcommand{\Hi}{\mathcal{H}}

\newcommand{\dagg}{\dagger}
\newcommand{\1}{\mathbbm{1}}

\newcommand{\lv}{\left \vert}
\newcommand{\rv}{\right \vert}
\newcommand{\la}{\left \langle}
\newcommand{\ra}{\right \rangle}
\newcommand{\ket}[1]{\lv #1 \ra}
\newcommand{\bra}[1]{\la #1 \rv}

\theorembodyfont{\rmfamily}
\newtheorem{Theorem}{{Theorem} }

\newtheorem{Lemma}{{Lemma}}
\newtheorem{Proof}{{Proof}}

\newcommand{\R}{\mathbbm{R}}

\newcommand{\id}{\mathbbm{1}}

\renewcommand{\tr}{{\rm Tr}\,}
\renewcommand{\det}{{\rm Det}\,}

\newcommand{\be}{\begin{equation}}
\newcommand{\ee}{\end{equation}}
\newcommand{\bea}{\begin{eqnarray}}
\newcommand{\eea}{\end{eqnarray}}

\newcommand{\eq}[1]{Eq.~(\ref{#1})}

\newcommand{\tra}[1]{{\rm Tr}\left[#1\right]}

\begin{document}

\title[Quantum benchmarks for teleportation and storage of squeezed states]
{Squeezing the limit: Quantum benchmarks for the teleportation and storage of squeezed states}

\date{14-08-2008}

\author{M. Owari,$^{1,2}$ M. B. Plenio,$^{1,2}$ E. S. Polzik,$^3$ A. Serafini,$^{4}$ and M. M. Wolf $^{3}$}

\address{$^1$ Institute for Mathematical Sciences, 53 Prince's Gate,
Imperial College London, London SW7 2PG, UK\\
$^2$ QOLS, Blackett Laboratory, Imperial College London, London SW7 2BW, UK}

\address{$^3$ Niels Bohr Institute, Copenhagen University, Blegdamsvej 17, DK- 2100
Copenhagen {\O}, Denmark}

\address{$^4$ Department of Physics \& Astronomy, University College London, 
Gower Street, London WC1E 6BT}

\ead{\mailto{mail}}
\begin{abstract}
We derive fidelity benchmarks for the quantum storage and teleportation of squeezed states
of continuous variable systems,
for input ensembles where the degree of squeezing $s$ is fixed, no information
about its orientation in phase space is given, and the distribution of phase space displacements
is a Gaussian.
In the limit where the latter becomes
flat, we prove analytically that the maximal classical achievable fidelity
(which is $1/2$ without squeezing, for $s=1$)
is given by $\sqrt{s}/(1+s)$,
vanishing when the degree of squeezing diverges.
For mixed states, as well as for general distributions of displacements,
we reduce the determination of the benchmarks to the solution of a finite-dimensional semidefinite program,
which yields accurate, certifiable bounds thanks to a rigorous analysis of the truncation error.
This approach may be easily adapted to more general ensembles of input states.
\end{abstract}

\vspace{2pc}

\submitto{\NJP}

\maketitle

\section{Introduction}
The storage and entanglement assisted teleportation of quantum states are two
of the central primitives of Quantum Information Science.
They have by now been accomplished with increasing precision in various experimental settings,
with outstanding examples in the `continuous variable' regime,
adopting light modes or collective spins of atomic ensembles to, respectively, carry and store
quantum information \cite{furusawa1,furusawa2,polzik1,polzik2}, or relying on motional atomic degrees of freedom \cite{atoms1,atoms2}.
The need to certify success in such experiments, and to justify the use of the term \emph{quantum}
in setups such as `quantum memories' and `quantum teleportations', requires theoretical benchmarks
which bound the performance of purely \emph{classical}
schemes \cite{BFK00,Hammerer,serafozzi,Adesso,calsamiglia}.
Here, ``classical'' refers to protocols where the quantum system is measured and later re-prepared
from information obtained in the measurement, which is in turn stored or transmitted by classical means.
The fact that there are limitations to such `measure-and-prepare' schemes
immediately follows from the no-cloning principle.
A precise notion of these limitations, however,
depends on the figure of merit (usually the \emph{fidelity}) as well as on the prior distribution,
{\em i.e.}, on the ensemble of quantum states to be stored or transmitted.

If, for a $d$-dimensional Hilbert space, the input ensemble is comprised of all the pure states
distributed according
to the Haar measure,
the average fidelity achievable by a classical scheme is $2/(d+1)$, dropping to zero
with increasing dimension $d$ \cite{Dbound,Dbound2}.
Clearly, for \emph{continuous variable systems} (where $d=\infty$)
not all pure states are experimentally accessible
(as they constitute a set with infinitely many real parameters and, in principle,
unbounded energy).
The typical continuous variable implementations via electro-magnetic field modes or collective fluctuations
in atomic ensembles favor Gaussian states of which coherent states are the simplest representatives.
It was proven in \cite{Hammerer} that the optimal  fidelity for classical schemes is $1/2$
if the input ensemble is made up of coherent states taken from a flat distribution in phase space.
This shows that such restrictions on the `alphabet' of input states
allows for classical schemes to achieve finite average fidelities, even though the dimension of
the Hilbert space is infinite.

The present work deals with input ensembles
including Gaussian {\em squeezed states},
for which one expects, in general, the constraint on classical schemes to become more and more severe
as the degree of squeezing is increased.
This study finds its motivation in recent and ongoing experimental attempts to teleport and store
squeezed quantum states \cite{singleteleport1,singleteleport2,squeezedmemory1,squeezedmemory2}:
we will provide a means for certifying success in experiments.

Although the tools we shall develop are suitable for more general applications,
encompassing non-Gaussian states or even finite dimensional systems, the focus of the
present paper will be Gaussian input ensembles with a fixed degree of squeezing $s$,
($s$ being the factor by which the variance of one of the two canonical quadratures is reduced
with respect to the vacuum level). This complements the results of Ref.~\cite{Adesso},
where $s$ is assumed to be unknown, and generalizes the results of Ref.~\cite{Hammerer}
on coherent states (special case $s=1$ in our notation).
In brief, we have been able to obtain the following results:
\begin{enumerate}
    \item\label{it1} For an ensemble of squeezed coherent states where, apart from the degree of squeezing,
    no a priori information is given about orientation and displacement,
    the maximal classical achievable fidelity is $\sqrt{s}/(1+s)$.
    The optimal measure-and-prepare scheme achieving this bound is
    realized by heterodyne detection followed by preparation of a coherent state.
\item For an ensemble of squeezed states as in (\ref{it1}), but where each state is subject to additive Gaussian noise,
analytical bounds are derived for the fidelity as well as for the maximal overlap achievable by classical schemes.
\item For ensembles with a random orientation (phase covariance) but arbitrary distribution of displacements
the maximal overlap is shown to be computable by means of a semidefinite program (SDP).
Together with a rigorous bound on the truncation error in Fock space this yields reliable benchmarks.
\item The SDP is applied to various cases of pure state ensembles
(with and without Gaussian displacement distribution) and mixed state ensembles
for parameter regimes relevant to present and future experiments.
\end{enumerate}

Further technical details can be found in the Appendices, where we present
a characterization of (time-reversible) covariant channels and a discussion of the Choi-matrix formalism
in infinite dimensions.

\section{Notation and conventions}\label{notation}

In this Section we briefly recapitulate some basic notation and useful
concepts. For a more detailed exposition the reader is referred to Refs.~\cite{eisertplenio,gaussianbook,Gausschannel}.

Throughout the paper, we will deal with bosonic systems of $n$ modes,
each of which is assigned to an infinite-dimensional Hilbert space
$\Hi=L_{2}(\R)$.
$\B(\Hi)$ and
${\mathfrak C}_1(\Hi)$ will denote the
set of bounded operators and the set of
 trace-class operators, respectively.
Let us arrange the $2n$ canonical operators in a vector
${R} = ({X}_1,{P}_1,\ldots,{X}_{n},{P}_{n})^{\sf T}$
such that $[{R}_{j},{R}_k] = i \sigma_{jk}\1$,
$\sigma = i\oplus_{j=1}^{n} \sigma_{y}$
being the anti-symmetric symplectic form
(while $\sigma_{y}$ is the two-dimensional $y$-Pauli matrix).

As is customary, the Weyl displacement operator $W_{\xi}$
will be defined as $W_{\xi}={\rm e}^{i\xi\cdot\sigma{R}}$,
for $\xi\in\R^{2n}$ so that $W_{\xi}W_{\eta}=W_{\eta}W_{\xi}{\rm e}^{-i\xi\cdot\sigma\eta}$.
The `characteristic function' $\chi_{{A}}(\xi)$ of an operator ${A}\in{\mathfrak C}_1(\Hi^{\otimes n})$
is defined as
\be
\chi_{A}(\xi) = \tra{W_{\xi}A} \; .
\ee
In turn, the operator $A$ is determined by its characteristic function
according to the Fourier-Weyl relation
\be
A = \left(\frac{1}{2\pi}\right)^{n} \int_{\R^{2n}}
\chi_{A}(\xi) W^{\dag}_{\xi} {\rm d}^{2n}\xi \; , \label{fourierweyl}
\ee
which leads to the useful Parseval relation for ${A}_1,{A}_2\in{\mathfrak C}_1(\Hi^{\otimes n})$:
\be
\tra{{A}_1^\dagger{A}_2} = \left(\frac{1}{2\pi}\right)^{n}
\int_{\R^{2n}} \overline{\chi_{A_1}(\xi)}\chi_{A_2}(\xi) {\rm d}^{2n}\xi \; .
\label{trace}
\ee
For a density operator $\rho$,
we  define a `covariance matrix' (CM) $\gamma_\rho$ with entries
$$
(\gamma_\rho)_{jk} = \tra{\{{R}_j,{R}_k\}\rho} - 2\tra{{R}_j\rho}\tra{{R}_k\rho}
$$
and a vector of first moments $d_{\rho}$ with entries $(d_{\rho})_j=\tra{{R}_j\rho}$.
A state $\rho$ is said to be Gaussian if its characteristic function $\chi_{\rho}$ is a Gaussian, reading
$$
\chi_{\rho} (\xi) = {\rm e}^{i\xi \cdot \sigma d_{\rho}- (\sigma \xi)
\cdot \gamma_{\rho} (\sigma\xi)/4} \; .
$$
The vacuum $\ket{0}\bra{0}$ is a Gaussian state with $\gamma_{\ket{0}\bra{0}}=\id$
and $d_{\ket{0}\bra{0}}=0$.

\section{Setup and figures of merit}


Our goal is to quantify the limitations for measure-and-prepare schemes $T$, {\em i.e.}, elements of the set ${\mathcal E}\ni T$  of \emph{entanglement-breaking channels} \cite{EB,HolevoEB}, when acting on an ensemble $\{\rho_\omega\}$ of input states characterized by a set of parameters $\omega$.
Consider for instance an ensemble of pure squeezed Gaussian states with CM
\be\label{eq:gammast}\gamma=\left(\begin{array}{cc}\cos\theta &\sin\theta\\-\sin\theta & \cos\theta
\end{array}\right)\left(\begin{array}{cc} s &0\\ 0 &1/s
\end{array}\right)\left(\begin{array}{cc}\cos\theta &-\sin\theta\\\sin\theta & \cos\theta
\end{array}\right),
\ee and displacement $\xi$ and thus $\omega=(s,\xi,\theta)$.
In order to fix a useful \emph{figure of merit} we have to choose a functional
$F(T(\rho_\omega),\rho_\omega)$ which
(i) measures, in some sense to be specified, the ability of $T$
to `preserve' the state $\rho_\omega$ when applied to it
and (ii) can be determined in experiments.
Based on this choice, we can then either quantify the \emph{worst case performance}
or the \emph{average case performance} of a channel $T$,
where the latter depends on an a priori distribution $q(\omega)$ over the  parameter space.
The corresponding benchmarks are then obtained by taking the supremum
over all $T\in\mathcal E$ leading to the definitions:
\bea F_0(T):=\inf_{\rho_\omega}\; F(T(\rho_\omega),\rho_\omega),\qquad\qquad F_0:=\sup_{T\in\mathcal E} F_0(T),\\
\overline{F}(T):=\int d\omega\; q(\omega) F(T(\rho_\omega),\rho_\omega),\qquad \overline{F}:=\sup_{T\in\mathcal E} \overline{F}(T).\eea
We will restrict attention to ensembles $\{\rho_\omega\}$ with a fixed degree of squeezing $s$, so that the benchmarks
will be functions of $s$ and we will occasionally write $\overline{F}(s),F_0(s)$ to emphasize this dependence.

As usual, for the ideal scenario of pure states $\rho_\omega$, we use the \emph{fidelity} $F(T(\rho_\omega),\rho_\omega)=\tr{[T(\rho_\omega)\rho_\omega}]$.
Clearly, in practice it is more realistic to assume that the initial pure states $\rho_\omega$
undergo a noisy channel $\mathcal N$ and become mixed before
entering a quantum memory or a teleportation scheme.
A possible option for $F$ would then be the ``Uhlmann fidelity'' \cite{U76} between two mixed states,
given by $f(\rho_1,\rho_2)=\tr\Big[\sqrt{\rho_1^{1/2}\rho_2\rho_1^{1/2}}\Big]^2$.
Major drawbacks of this choice are that its non-linearity leads to a very involved theoretical optimization
and that such a quantity is exceedingly difficult to measure in experiments
(without invoking a plethora of extra assumptions).
We will present later on an analytical result adopting Uhlmann fidelity,
but we will mainly follow a different route
and use instead the \emph{overlap}
\be F(T(\rho_\omega),\rho_\omega)=\tr\big[T\big({\mathcal N}(\rho_\omega)\big)\rho_\omega\big].\label{eq:overlap}\ee
This quantity is easier to determine in experiments
($\rho_\omega$ may be interpreted as an observable) and
by definition $F_0$ and $\overline{F}$ are proper benchmarks in the sense that beating their values means outperforming any classical scheme. Strictly speaking, of course, this overlap measures how close the output of $T$ is to the initial noiseless state rather than to the input.

The derivation of the benchmarks simplifies considerably if the probability measure $q$ is invariant
with respect to a symmetry group $G$, {\em i.e.}, if  $q(\omega)=q\big(g(\omega)\big)$ for all $g\in G$ and $\rho_{g(\omega)}=U_g\rho_\omega U_g^\dagger$
for some unitary representation $U_g$ of $G$.
A standard argument \cite{Dbound,OleCloning} then implies that, w.l.o.g.,
the channels $T$ in the optimization of $F_0$ and $\overline{F}$
can be taken covariant with respect to $G$, in the sense that, for all density operators $\rho$ and all $g\in G$, one has
\be U_g T(\rho)U_g^\dagger =T\big(U_g\rho U_g^\dagger\big). \ee
 For a compact symmetry group (with Haar measure $dg$) the argument is straight forward since we can replace every single $T$ by a covariant counterpart
\be \tilde{T}(\rho):= \int dg\; U_g^\dagger T\big(U_g\rho U_g^\dagger\big)U_g,\ee
which performs at least as well as $T$ if the chosen functional $F$ is concave,
as is the case for all the instances we discussed above.
Moreover, if $T$ is an element of a convex set closed under the group action,
like the set of entanglement breaking channels, then so is $\tilde{T}$.
If the orbit of $G$ covers the entire parameter space then  $F(\tilde{T}(\rho_\omega),\rho_\omega)$
becomes state independent and we find $F_0=\overline{F}$, {\em i.e.},
the average case and the worst case performance become the same.

The argument becomes more subtle if $G$ is not compact,
as in the case where no a priori information is given about the displacement $\xi$.
In this case $F_0$ is still well defined but already $\overline{F}$
has to be discussed more carefully, since it formally requires an average over a `flat' distribution in phase space.
Nevertheless, an analogous argument goes through and we can w.l.o.g.~take $T$ to be
\emph{phase space covariant}, {\em i.e.}, for all $\xi$ and density operators $\rho$ \be W_\xi T(\rho)W_\xi^\dagger =T\big(W_\xi\rho W_\xi^\dagger\big). \ee  Clearly, due to non-compactness one cannot make the averaging procedure explicit, but has rather to invoke an invariant mean whose existence is guaranteed by the axiom of choice. For a more formal discussion of these matters see \cite{OleDiplom,WernerUnc}.
\section{Analytical benchmarks}


\subsection{Classical benchmark for pure squeezed states
under uniform rotations and displacements}\label{infinite_pure}

Let us now focus on the specific case of an initial single-mode squeezed state $\rho_{s}$  with CM
$$
\gamma_{\rho_s} = \left(\begin{array}{cc}
s & 0 \\
0 & 1/s
\end{array}
\right)  \, ,
$$
which is uniformly displaced and rotated in phase space.
The input ensemble is given by $\rho_\omega=W_\xi U_\theta\rho_s U_\theta^\dagger W_\xi^\dagger$,
where $U_{\theta}=\exp[i\theta \hat{n}]$ denotes a phase space rotation
(while $W_{\xi}$ is a displacement operator, defined in Section \ref{notation}), and
no a priori information is given about $\theta \in [0,2\pi]$ and $\xi \in \R^{2}$,
whereas the degree of squeezing $s$ is fixed and known a priori.

In order to determine the best possible fidelity achievable by a
measure-and-prepare scheme on such an ensemble, we will first prove an upper bound by
enlarging the set of allowed channels to the set ${\mathcal T}\supseteq\mathcal E$ of
time-reversible channels. That is, to all the channels which remain valid
quantum operations when concatenated with time-reversal
(this includes, for instance, all measure-and-prepare schemes which
are assisted by PPT bound-entanglement shared between sender and
receiver). In a second step we will show that the obtained bound is
tight, even within the set ${\mathcal E}$,
as it turns out to be achievable by heterodyne measurement
followed by preparation of a coherent state.

\begin{Theorem}[Benchmark---pure states]\label{bench_pure}
{\em Let $\cal T$ be either the set of all measure-and-prepare schemes
or the larger set of time-reversible channels. Within these sets,
the maximal achievable fidelity for the teleportation or storage
of squeezed coherent states with squeezing $s$
and subject to uniformly distributed rotations and displacements in phase space is given by
\be
\overline{F}(s)={F}_0(s) = \sup_{T\in\cal T}\; \inf_{\xi,\theta} \;
\Tr\left[\rho_\omega
T(\rho_\omega)\right] =
\frac{\sqrt{s}}{1+s}.\label{eq:sbound}
\ee}
\end{Theorem}
\begin{Proof}
{We first prove the upper bound.
As already indicated above
we can restrict ourselves to the set $\tilde{\mathcal T}$ of \emph{phase space covariant}
channels satisfying
\be
\tilde{T}(W_\xi A W_\xi^\dagg)=W_\xi \tilde{T}(A)W_\xi^\dagg,
\ee
for all Weyl operators $W_\xi,\;\xi\in\mathbb{R}^2$
and all $A\in {\mathfrak C}_1 \left(L_{2}(\R)\right)$.
This leads to $F_0=\overline{F}$ and lifts the need
for the minimization over displacements \cite{OleDiplom,WernerUnc}.
That is, we are left to determine
\be\label{eq:imohr}
 \sup_{\tilde{T}\in\tilde{\cal T}} \inf_{\theta\in[0,2\pi]}
\Tr\left[U_{\theta}
\rho_{s} U^{\dag}_{\theta}\tilde{T}(U_{\theta}
\rho_{s} U^{\dag}_{\theta})\right]
\ee
(in other words, the infimum is only taken over all squeezed coherent states
centred at the origin, {\em i.e.}, with vanishing first moments).

Furthermore, time-reversible phase space covariant channels have
a particular simple form in the Heisenberg picture, where we will
denote the map under consideration by $\tilde{T}^*$.
By Lemma \ref{lem:timerev} (see \ref{phspcovariant})
they act on Weyl operators as
\be
\tilde{T}^*(W_\xi)=t(\xi)W_\xi
\ee
where
\be
t(\xi)=\Tr\left[\tau W_{\sqrt2 \xi}\right]
\ee
for some density matrix $\tau$. Conversely, every such $\tau$ yields an
admissible phase space covariant channel $\tilde{T}$.
This allows us to recast the
optimization over channels into one over density operators, as follows.

Exploiting the Parseval relation (\ref{trace}) back and forth, we get
for any density matrix $\rho$ whose characteristic function is a centred Gaussian:
\bea
\tra{\rho \tilde{T}(\rho)} &=& \frac{1}{2\pi} \int {\rm d}^{2}\xi
\tra{\rho W_{\xi}}^2 \tra{\tau W_{\sqrt{2}\xi}} \nonumber \\
&=&
\frac{1}{2\pi} \int {\rm d}^{2}\xi
\tra{\rho W_{\sqrt{2}\xi}} \tra{\tau W_{\sqrt{2}\xi}} = \frac12 \tra{\rho\tau}
\eea
(where we took advantage of the fact that the characteristic function of centred Gaussian
states is real and has a purely quadratic dependence on $\xi$).

To compute Eq.(\ref{eq:imohr}) it is now convenient to reintroduce an average over rotations
(instead of an infimum, which gives the same value due to the optimization over $T$),
and write Eq.(\ref{eq:imohr}) as:
\be
\hspace*{-1.6cm}
 \sup_{\tau} \tra{\frac{1}{4\pi}\left(\int_{0}^{2\pi}
U_{\theta}
\rho_{s} U^{\dag}_{\theta} {\rm d}\theta\;\right) \tau} =
\left|\left|\frac{1}{4\pi}\int_{0}^{2\pi}
U_{\theta}
\rho_{s} U^{\dag}_{\theta} {\rm d}\theta \right|\right|_{\infty} \; .\label{eq:suptauinfrho}
\ee
This operator
norm (largest eigenvalue) can be promptly determined since averaging over rotations
just sets all off-diagonal elements in Fock basis to zero, so that one simply has to resort
to the expression of a squeezed vacuum state in Fock basis, given by
$$\sqrt{\frac{2\sqrt{s}}{1+s}}\sum_{n=0}^\infty \frac{\sqrt{(2n)!}}{n!}\left(\frac{s-1}{2s+2}\right)^n |2n\rangle. $$
As the largest diagonal entry is the vacuum component we finally obtain
$$\overline{F}(s) \leq \frac12\bra{0}\rho_s\ket{0} = \frac{\sqrt{s}}{1+s} \; ,$$
where the r.h.s. is attained within the
set of time-reversible channels.

To show that this upper bound is actually tight for measure-and-prepare schemes as well,
we will now explicitly point out that
a specific, simple measure-and-prepare scheme attains the bound \cite{bound_note}.
Consider a heterodyne measurement, {\em i.e.}, a POVM
$\{W_\xi|0\rangle\langle 0| W_\xi^\dagg/2\pi\}$, where the outcome $\xi$ is followed by the
preparation of a coherent state $W_\xi|0\rangle$,
so that the whole map ${ H} $
reads
\be
{ H}(\rho) = \frac1{2\pi} \int {\rm d}^2 \xi \bra{0}W^{\dag}_{\xi} \rho W_{\xi}\ket{0}
W_{\xi}\ket{0}\bra{0}W^{\dag}_{\xi} \label{heterodyne}
\ee
This process
acts on covariance matrices as $\gamma\mapsto\gamma+2\cdot\1$
(while leaving first moments unaffected), so that one can
easily determine the input-output fidelity for any state in our ensemble using again Parseval's relation (\ref{trace}) and solving the resulting Gaussian integrals \cite{Scutaru}
\be
\tra{\rho_\omega{ H}(\rho_\omega)}=
2\Big[\det{\Big(\gamma_{\rho_s}+(\gamma_{\rho_s}+2\cdot \1)}\Big)\Big]^{-1/2}=\frac{\sqrt{s}}{1+s},
\ee
which is independent of $\xi$ and $\theta$ and achieves the above bound.}
\hfill $\square$
\end{Proof}

\begin{figure}[ttt]
\begin{center}
\epsfig{file=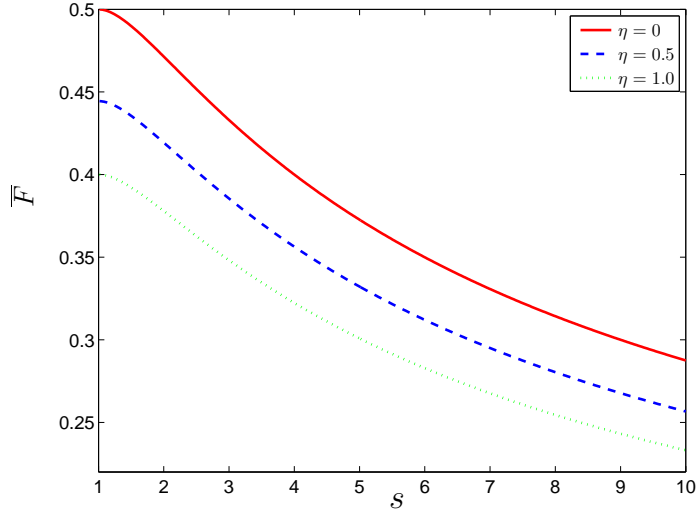,angle=0,width=0.7\linewidth}
\vspace*{-0.4cm}
\end{center}
\caption{Fidelity benchmarks $\overline{F}(s,\eta)$ for
measure-and-prepare schemes on ensembles with squeezing $s$,
flatly distributed displacements and random phase space orientations,
as a function of the squeezing $s$ (before additive noise is applied).
The continuous (red) curve refers to pure states ($\eta=0$)
while the dashed (blue) and dotted (green) curves refer to mixed states, for $\eta=0.5$
and $\eta=1$ respectively.
Notice that, as one should expect, the pure state case is an upper bound for mixed states for
the figure of merit $\overline{F}$. As stated in Theorem \ref{bench_uhlmann},
the pure states case is also an upper bound for the Uhlmann fidelity benchmark of
such mixed states.} \label{figure:set}
\end{figure}

Note that the optimal strategy (heterodyning followed by generation of coherent states)
does not depend on the degree of squeezing and is in fact the very same strategy which
is optimal for a flat distribution of coherent states \cite{BFK00}
(for coherent states, such optimality extends to rotationally invariant, Gaussian distributions
of displacements as well \cite{Hammerer}).
The optimal fidelity as a function of the squeezing $s$ is plotted in Fig.~\ref{figure:set}.

This analytical result, though obtained for an ideal case, fully highlights the importance of 
radomising the input phase in practical instances. 
In fact, for a `flat' distribution of displacements but fixed phase, the benchmark 
can be promptly inferred from the coherent states' case, and is just $1/2 \ge \sqrt{s}/(1+s)$,
regardless of the degree of squeezing $s$ assumed. 
Randomising the input phase appears to be very helpful 
to provably reach the quantum regime in experiments (and, not surprisingly, 
the advantage granted by random rotations becomes more and more relevant 
with increasing squeezing): for pure states and degrees of squeezing within experimental reach
($s\lesssim10$), the difference is around $0.2$ and might easily 
turn out to be crucial for experimental success.
As we will see later on (see Sec.~\ref{section numerical calculations}), 
and again not surprisingly, the same holds true for the randomisation of displacements.

\subsection{Classical benchmark for mixed squeezed states
under uniform rotations and displacements}\label{infinite_mixed}

Let us now consider the case of an input ensemble of mixed
Gaussian states, derived from the initial pure states $\rho_\omega$, $\omega=(s,\xi,\theta)$
by the application of a noisy Gaussian channel $\mathcal N$,
adding classical Gaussian noise with variance $\eta$, and
thus acting on covariance matrices as $\gamma\mapsto\gamma+\eta \1$
while leaving first moments invariant \cite{Gausschannel}.
$\mathcal N$ can be understood as the application of random displacements
to the input state, distributed according to a Gaussian with variance $\eta$.
We will keep referring to $s$ as squeezing parameter (here, prior to the action of noise)
and will refer to $\eta$ as to the `additive noise'.

As pointed out earlier we will use the overlap with $\rho_\omega$ as figure of merit: $$F_0(s,\eta)=\sup_{T\in\mathcal E}\;\inf_{\xi,\theta}\;\tr\left[T\Big({\mathcal N}(\rho_\omega)\Big)\rho_\omega\right] \, ,$$
where, due to the symmetry (which is preserved by $\mathcal N$), one has again $F_0=\overline{F}$.
For the same reason we can again restrict to phase space covariant channels and proceed along the lines of
Theorem \ref{bench_pure}:

\begin{Theorem}[Benchmark---mixed states]\label{bench_mixed}
{\em Let $\cal T$ be either the set of all measure-and-prepare schemes
or the set of time-reversible channels. Within these sets
the maximal achievable overlap for the teleportation or storage
of mixed squeezed coherent states with squeezing $s$ and
additive noise $\eta$ is given by
\bea
\overline{F} (s,\eta) &=& \sup_{T\in\cal T} \;\inf_{\xi,\theta}\;\tr\left[T\Big({\mathcal N}(\rho_\omega)\Big)\rho_\omega\right] \nonumber \\
&=&
\left[\Big(1+\frac{\eta}2+\frac{1}{s}\Big)\Big(1+\frac{\eta}2+{s}\Big)\right]^{-1/2} \, .
\label{eq:Bmixedoverlap}
\eea}
\end{Theorem}
\begin{Proof}
{As for Theorem \ref{bench_pure}, we restrict to phase space covariant channels $\tilde{T}$
parameterized by a density operator $\tau$ and obtain in an analogous way
\be \hspace*{-1cm}
\tr\left[\tilde{T}\Big({\mathcal N}(\rho_\omega)\Big)\rho_\omega\right]
= \frac{1}{2\pi}\int {\rm d^2}\,\xi\;
\tra{\tau
W_{\sqrt{2}\xi}} e^{-\frac12 \xi\cdot\Gamma\xi}
=\frac12\tra{\tau\rho^{\prime}} \; ,
\ee
where $\rho^{\prime}$ is now a centred
Gaussian state with covariance matrix $\Gamma=\gamma_{\rho_s}+\eta/2$.
Once again, the supremum over $\tau$ and thus over all time-reversible channels  can be calculated by considering the state $\rho^{\prime}$ in the number basis, yielding
\bea \hspace*{-2.0cm}
\sup_{\tau}\frac1{4\pi}
\tra{\left(\int_{0}^{2\pi} {\rm d}\theta U_{\theta}\rho^{\prime}U^{\dag}_{\theta}\right)\tau}&=&
\left|\left|\frac1{4\pi}\int_{0}^{2\pi}d\phi\;
U_\phi \rho^{\prime} U_\phi^\dagger\right|\right|_{\infty}\nonumber\\
=\ \max_n \frac12 \langle n|\rho^{\prime}|n\rangle&=&\frac12 \langle
0|\rho^{\prime}|0\rangle
=\left[\Big(1+\frac{\eta}2+\frac{1}{s}\Big)\Big(1+\frac{\eta}2+{s}\Big)\right]^{-1/2},\nonumber
\eea
which is
Eq.(\ref{eq:Bmixedoverlap}) for time-reversible channels.
Again, it can be shown by direct evaluation that the heterodyne strategy
described by the map $H$ of \eq{heterodyne} attains this bound,
which completes the proof.}
\hfill$\square$
\end{Proof}

Notice that the overlap achieved by the ideal quantum channel (identity)
is $\left[\Big(\frac{\eta}2+\frac{1}{s}\Big)\Big(\frac{\eta}2+{s}\Big)\right]^{-1/2}$.
Also, $\overline{F}(s,0)=\overline{F}(s)$, as it should be
since both the input ensemble and the figure of merit coincide with
those of the preceding section in the noiseless case ($\eta=0$).
Clearly, the optimal classical performance degrades with increasing noise
(this is obviously the case for quantum strategies as well).
Examples of such benchmarks for mixed states are displayed in Fig.~\ref{figure:set}.

For the same ensemble of mixed state, we shall also present a further analytical benchmark
in the form of an upper bound,
this time adopting the Uhlmann fidelity $f(T(\rho),\rho)$ as figure of merit:
\begin{Theorem}[Benchmark---mixed state fidelity]\label{bench_uhlmann}
{\em Let $\cal T$ be either the set of all measure-and-prepare schemes
or the set of time-reversible channels. Within these sets
the maximal achievable worst case (or average) Uhlmann fidelity for the teleportation or storage
of mixed squeezed coherent states with squeezing $s$ and
arbitrary additive noise channel $\mathcal N$ is bounded as follows
\be
\sup _{T \in {\cal E}} \inf_{\xi,\theta}f\big(T\left({\mathcal N}(\rho_\omega)\right),{\mathcal N}(\rho_{\omega})\big)\leq \frac{\sqrt{s}}{1+s} \quad \forall\, \eta\in\R \, .
\ee}
\end{Theorem}
\begin{Proof}{Concavity of the fidelity allows us again to restrict to
phase space covariant channels. All these channels commute with
channels of the form ${\mathcal N}^*(W_\xi)=W_\xi\exp[-||\xi||^2\eta/4]$
(${\mathcal N}^*$ standing for channel ${\mathcal N}$ in Heisenberg picture). The
result follows then from the
contractivity of cp-maps with respect to the fidelity, together
with the pure state result of Theorem \ref{bench_pure} [Eq.(\ref{eq:sbound})]. }
\hfill$\square$
\end{Proof}

\section{Quantum benchmarks derived by Semidefinite-Programming}
\subsection{Problem settings} \label{section: problem settings}
So far, we treated  input ensembles of squeezed states which are displaced in phase space according to a `flat' distribution.
Needless to say, this is an idealization as one cannot implement an arbitrarily large displacement in practice.

To be more realistic we have to treat an input ensemble of squeezed states whose first moments are essentially contained
in a finite region of  phase space, {\em e.g.}, due to a sufficiently rapidly decaying probability  distribution as in \cite{BFK00, Hammerer}.

This section will deal with this scenario by resorting to numerical means, as a purely analytical
treatment appears to be far too involved for finite, non-flat distributions of displacements
(where the restrictions to phase space covariant channels is no longer optimal).
We will show that the problem of computing  benchmarks of the desired kind  can be cast into a \emph{semidefinite program} (SDP). As such, it comes with a guarantee of computing the correct value,
as SDPs come in pairs of a \emph{primal} and a \emph{dual} problem which yield converging upper and lower bounds to the sought solution \cite{Boyd}.
As the original SDP is in an infinite dimensional space, truncation will be necessary and will induce errors.
However, we will provide a rigorous bound to the truncation errors so that the finally derived benchmarks
are reliable
and rigorous, constituting upper bounds to the actual optimal classical figures of merit.

The figure of merit we use is again the average overlap $$\overline{F}=\sup_{T\in\mathcal E}\int d\omega\;q(\omega)\tra{T\Big({\mathcal{N}}(\rho_\omega)\Big)\rho_\omega}, $$
where $\mathcal N$ is a noisy channel that describes the noise suffered by $\rho_\omega$
before entering the storage or teleportation device,
which for our purposes should be
\begin{enumerate}
\item a channel which allows for the computation of the matrix elements in the Fock basis
$\langle k|{\mathcal N}(\rho_\omega)|l\rangle$, {\em e.g.}, a Gaussian channel \cite{Gausschannel} 
\item rotationally covariant, {\em i.e.}, $\mathcal{N}\left ( U_{\theta} \rho U_{\theta}^{\dagger} \right )=
U_{\theta} \mathcal{N}\left (  \rho  \right ) U_{\theta}^{\dagger}$ for all  $\rho,\theta$.
\end{enumerate}
Fortunately, channels representing attenuation, amplification, and thermal noise
are all of this type,
the attenuation channel being probably the most relevant here, as it models losses of photons/excitations
which are the dominant decoherence process in the practical realisations we have in mind, involving
traveling waves of light.
Thus we will adopt it in the following, denoting it by $\mathcal{N}_{\lambda}$. The parameter $\lambda \in [0,1]$ represents the transmitivity (in intensity)
such that the channel acts on, respectively,
 covariant matrix $\gamma _{\rho}$ and vector of first moments
$d_{\rho}$
as: $\gamma _{\rho} \mapsto
\lambda \gamma _{\rho} + (1-\lambda) \1 $ and $d_{\rho} \mapsto \sqrt{\lambda}d_{\rho}$.
Another crucial assumption for our method is that the input distribution $q$ is uniform in the angle $\theta$.
Moreover, we will consider ensembles with constant degree of squeezing $s$
(although this is not necessary for the method).
Hence $q$ can be considered a probability distribution which depends only on the displacement $\xi$ and therefore
\be\label{abstract setting}
 \hspace{-1cm}\overline{F} =
\sup _{T \in \mathcal{E}} \int _{\theta \in [0,2\pi]} \int _{\xi \in \mathbb{R}^2}
  q(\xi) \tr \left[ T \Big(\mathcal{N}_\lambda(U_\theta \rho_{s,\xi} U_\theta^{\dagger} )\Big)
\Big( U_{\theta} \rho_{s,\xi} U_{\theta}^{\dagger} \Big) \right] \frac{d\theta}{2\pi} d\xi,
\ee
where $\rho_{s,\xi}$ is a pure squeezed  states with degree of squeezing $s$ and displacement $\xi$.
Note that $\overline{F}$ in this way becomes a functional of $\lambda, s$ and of the distribution $q$.

\subsection{Reduction to a finite dimensional semidefinite program} \label{reduction of a problem}
In this subsection, we will see how we can reduce the quantum benchmark $\overline{F}$ in Eq.(\ref{abstract setting})
to a finite dimensional SDP.
We first show that Eq.(\ref{abstract setting}) can be reduced
to an infinite dimensional SDP problem.
To this end, we use a simple correspondence between the set of all entanglement breaking channels
and the set of  bipartite separable positive operators \cite{Rains} which is
 nothing but the Choi-Jamiolkowski isomorphism \cite{Choi Jamiolkowski}, albeit for an infinite dimensional system (see \ref{proof of isomorphism} for a proof):
\begin{Theorem}[Choi-Jamiolkowski]\label{theorem isomorphism}
{\em Suppose $\B (\Hi)$ is the space of all bounded operators and $\mathfrak{C}_1(\Hi)$
is the space of all trace class operators on a separable Hilbert space $\Hi$.
Then, for all entanglement breaking channels $\Psi$ on $\Hi$,
there exist a unique separable positive bounded operator $\Omega(\Psi)$ on $\Hi \otimes \Hi$ such that $Tr_B (\Omega(\Psi)) = \1_A$ and
\begin{equation}\label{isomorphism}
\Tr(B\Psi (A)) = \Tr (\Omega(\Psi) A\otimes B)
\end{equation}
for all $A \in \mathfrak{C}_1(\Hi)$ and $B \in \B(\Hi)$. Conversely,
for a separable positive bounded operator $\Omega$  on $\Hi \otimes \Hi$ satisfying  $Tr_B (\Omega) = \1_A$,
there exists a unique channel $\Psi(\Omega)$ such that it is entanglement breaking and satisfies Eq.(\ref{isomorphism}).}
\end{Theorem}

By means of the above theorem, we immediately derive an upper bound
to the quantity $\overline{F}$ of Eq.(\ref{abstract setting}) in an infinite dimensional SDP form by enlarging the set of positive separable operators (denoted by ${\rm Sep}$) to the set of positive operators with positive partial transpose (PPT) $\Omega^{\Gamma} \ge 0$:
\begin{eqnarray}\label{infinite SDP}
\overline{F} &=& \sup _{\Omega \in \B (\Hi \otimes \Hi)} \left \{ \Tr (\Omega \eta)
\Big|
  \Omega \in {\rm Sep }, \Tr _B \Omega = \1_A  \right \}
\nonumber \\
&\le& \sup _{\Omega \in \B (\Hi \otimes \Hi)} \left \{ \Tr (\Omega \eta)
\Big|
 \Omega \ge 0, \Omega ^{\Gamma} \ge 0, \Tr _B \Omega = \1_A  \right \}.
\end{eqnarray}
In the above equations $\eta$ is a state on $\Hi \otimes \Hi$ defined by
\be\label{eq: definition of eta}
\fl \eta  \stackrel{\rm def}{=}  \int _{\theta \in [0,2\pi]} \int _{\xi \in  \mathbb{R}^2}
  q(\xi) \ U_{\theta} {\mathcal N}_\lambda(\rho_{s,\xi}) U_{\theta}^{\dagger}
\otimes
\left ( U_{\theta} \rho _{s,\xi} U_{\theta}^{\dagger} \right )  \frac{d\theta}{2\pi}\; d\xi,
\ee
where we exploited the rotational covariance of $\mathcal{N}_{\lambda}$.

Here, we should remark that, if $\lambda=1$ and $q(\xi)= \frac{\alpha}{\pi}\exp [-\alpha ||\xi||^2]$ for $\alpha \ge 0$,
the inequality in Eq.(\ref{infinite SDP}) is actually an equality.
This is because, under such assumptions, we can choose an optimal $\Omega$ to be Gaussian \cite{Hammerer} and the
PPT condition $\Omega^{\Gamma} \ge 0$ is necessary and sufficient for a two-mode Gaussian state $\Omega$ to be separable \cite{simon00,WW01}.

In the remaining part of this subsection we will transform the above infinite dimensional SDP into a finite dimensional SDP.
We will need the following statement:
\begin{Lemma}[Operator norm bound]\label{operator norm bound}
If a positive separable operator  $\Omega \in \B(\Hi \otimes \Hi)$ satisfies $\Tr _B\Omega = \1_A $,
then, $\Omega$ also satisfies  $\| \Omega \|_{\infty} \le 1$, where $\| \cdot \|_{\infty}$ is the operator norm.
\end{Lemma}
\begin{Proof}
From the proof of Theorem \ref{theorem isomorphism}, $\Omega$ can be written as
$\Omega(\Psi) = \int _{\cal X} M(dx) \otimes \sigma(x)$  by using a POVM $\{ M(dx)\}$
and a set of states $\{ \sigma(x) \}$.
Then, for any normalized state $\ket{\Phi}$ on $\Hi \otimes \Hi$, we can bound $\bra{\Phi}\Omega\ket{\Phi}$ as follows:
\begin{eqnarray*}
\bra{\Phi} \Omega \ket{\Phi} &=& \int _{\cal X} \Tr (M(dx)\otimes \sigma(x) \ket{\Phi}\bra{\Phi} ) \\
 &\le& \int_{\cal X} \Tr (M(dx) \rho _A) = 1,
\end{eqnarray*}
where $\rho _A$ is defined as $\Tr _B(\ket{\Phi} \bra{\Phi})$ and we used that $\sigma(x)\leq\1$ together with $M({\cal X})=\1$.
Therefore, we have $\| \Omega \|_{\infty} = \sup _{ \Phi } \bra{\Phi}\Omega\ket{\Phi} \le 1$.
\hfill $\square$
\end{Proof}
Notice that, even if $\| \Omega \|_{\infty} \le 1$ was shown for the corresponding operator of an entanglement breaking channel,
this fact may not be true for other channels (in general
 we can only show that the geometric measure \cite{PV07}
$G(\Omega (T)) \stackrel{\rm def}{=} \sup_{\sigma \in Sep} \Tr \left ( \Omega (T) \sigma \right )\le 1$).

Before we derive a finite SDP problem from Eq.(\ref{infinite SDP}),
we observe the following crucial fact:
By means of the action of the group integral over $\{ U_{\theta} \otimes U_{\theta} \}_{\theta \in [0, 2\pi]}$,
$\eta$ is block diagonalized as
\begin{equation}\label{eq: eta block diagonal}
\eta = \sum _{c=0}^{\infty} Q_c\left [ \int _{\xi \in \mathbb{R}^2}
  q(\xi) \ \mathcal{N}_{\lambda}( \rho_{s,\xi})  \otimes  \rho _{s,\xi} \; d\xi \right ] Q_c,
\end{equation}
where $Q_c \stackrel{\rm def}{=} \sum _{k+l =c} \ket{k}\bra{k}\otimes \ket{l}\bra{l}$ is a finite dimensional projector.
Thus, defining  $P_c \stackrel{\rm def}{=} \sum _{i=0}^c Q_i$,
we obtain $\eta-P_c \eta P_c  = \sum _{i=c+1}^{\infty} Q_i \eta Q_i  \ge 0$ for all $c$.
We arrive at the finite SDP upper bound now as follows:
Suppose a separable positive $\Omega$ satisfies $\Tr _B \Omega =\1_A$.
Then,
\begin{eqnarray}\label{inequality one}
\Tr (\Omega \eta) &=& \Tr (\Omega P_c\eta P_c) + \Tr (\Omega(\eta-P_c \eta P_c)) \nonumber \nonumber \\
&\le& \Tr (\Omega P_c\eta P_c) + \| \Omega \|_{\infty} \Tr\big(\eta-P_c \eta P_c \big) \nonumber \\
&\le& \Tr (\Omega P_c \eta P_c) + \Tr (\eta-P_c \eta P_c) \nonumber \\
&=& \Tr (\Omega P_c \eta P_c) - \Tr (P_c \eta P_c) +1,
\end{eqnarray}
where we used the positivity of $\eta-P_c \eta P_c $ and Lemma \ref{operator norm bound}.

Finally, by means of Eq.(\ref{inequality one}), we can derive
an upper bound to the first line of Eq.(\ref{infinite SDP}).
Denoting with $R_c = \left ( \sum _{i=0}^c \ket{i}\bra{i} \right) \otimes \left
( \sum _{i=0}^c \ket{i}\bra{i} \right)$ the projection onto the
subspace with a photon number smaller than $c$ in each mode,
with support ${\rm supp}R_c$, we obtain
\begin{eqnarray}\label{finite SDP}
\fl \overline{F}
&\le &  \sup _{\Omega \in \B \left (\Hi  \right )  } \left \{ \Tr (\Omega P_c\eta P_c)
\Big| \Omega \in {\rm Sep }, \Tr _B \Omega = \1_A  \right \} + 1- \Tr (P_c\eta P_c ) \nonumber \\
\fl &= &  \sup _{\Omega \in \B \left (\Hi \right )  } \left \{ \Tr (R_c \Omega R_c P_c\eta P_c)
\Big| \Omega \in {\rm Sep }, \Tr _B \Omega = \1_A  \right \} + 1- \Tr (P_c\eta P_c ) \nonumber \\
\fl &\le &  \sup _{\Omega \in \B \left ({\rm supp}R_c  \right )  } \left \{ \Tr (\Omega P_c\eta P_c)
\Big| \Omega \in {\rm Sep }, \Tr _B \Omega \le \1_A  \right \} + 1- \Tr (P_c\eta P_c ) \nonumber \\
\fl &\le & \sup _{\Omega \in \B \left ({\rm supp}R_c  \right )  } \left \{ \Tr ( \Omega  P_c\eta P_c)
\Big| \Omega \ge 0, \Omega ^{\Gamma} \ge 0, \Tr _B \Omega \le \1_A  \right \} + 1- \Tr (P_c\eta P_c ),
\end{eqnarray}
The above upper bound now only involves a finite dimensional SDP, as
the original infinite dimensional $\Omega$ has been replaced by
the finite dimensional $R_c\Omega R_c$, with the same type of constraints
(note that $R_c$ was introduced to preserve the separability of the operator).
Similarly, the term $\Tr (P_c\eta P_c )$ is just a trace of a finite dimensional matrix, so that
we can numerically compute every term of the last formula of Eq.(\ref{finite SDP}) for as large a truncation parameter $c$
as our computer allows.
Thus, Eq.(\ref{finite SDP}) enables us to efficiently calculate an upper bound for the quantum benchmark $\overline{F}$
for any probability density $q(\xi)$, that is, ultimately, {\em for any rotationally-invariant input ensemble}.
Moreover, since in the limit of large $c$ we obtain the second formula of Eq.(\ref{infinite SDP}),
we can expect that this upper bound reaches the exact value for this bound for sufficiently large $c$.
Finally, we rephrase Eq.(\ref{finite SDP}) in the form of a theorem:
\begin{Theorem}[Benchmark---SDP]
{\em With the above definitions,
for any  probability density $q(\xi)$ on $ \mathbb{R}^2$
and rotationally covariant noise channel $\mathcal N$, we have
\begin{eqnarray}\label{finite SDP 2}
\fl\overline{F} \le \sup _{\Omega \in \B \left ({\rm supp}R_c
\right )  } \left \{ \Tr (\Omega  P_c\eta P_c) \Big| \Omega \ge 0,
\Omega ^{\Gamma} \ge 0, \Tr _B \Omega \le \1_A  \right \}
\quad + 1- \Tr (P_c\eta P_c ) \, . \nonumber\\ \hspace*{1cm}
\end{eqnarray} }
\end{Theorem}


\subsection{Results of numerical calculations} \label{section numerical calculations}
In this subsection, we present results of numerical calculations which were produced
using inequality (\ref{finite SDP 2}).
In order to reduce the memory requirements,
we further exploit the block diagonal structure of $\eta$ in Eq.(\ref{eq: eta block diagonal})
and impose $\sum _{j=0}^{2c} Q_j \Omega Q_j = \Omega$ for the implementation.
The evaluation of the bound then splits into two terms:
\begin{eqnarray*}
 \hspace{-1.4cm}\overline{F }_{finite}\stackrel{\rm def}{=} \sup _{\Omega \in \B \left ({\rm supp}R_c  \right )  } \left \{ \Tr (\Omega P_c\eta P_c)
\Big| \Omega \ge 0, \Omega ^{\Gamma} \ge 0,
\Tr _B \Omega \le \1_A \right \}, \\
 \hspace{-1.4cm}\epsilon_{error} \stackrel{\rm def}{=}  1- \Tr (P_c\eta P_c ). \\
\end{eqnarray*}
Denoting the r.h.s. of (\ref{finite SDP 2}) by $\overline{F}_{infinite}$ we have
\begin{equation}\label{eq:F _infinite}
\overline{F}_{infinite} = \overline{F }_{finite} + \epsilon_{error}.
\end{equation}

\begin{figure}[ttt]
\begin{center}
\epsfig{file=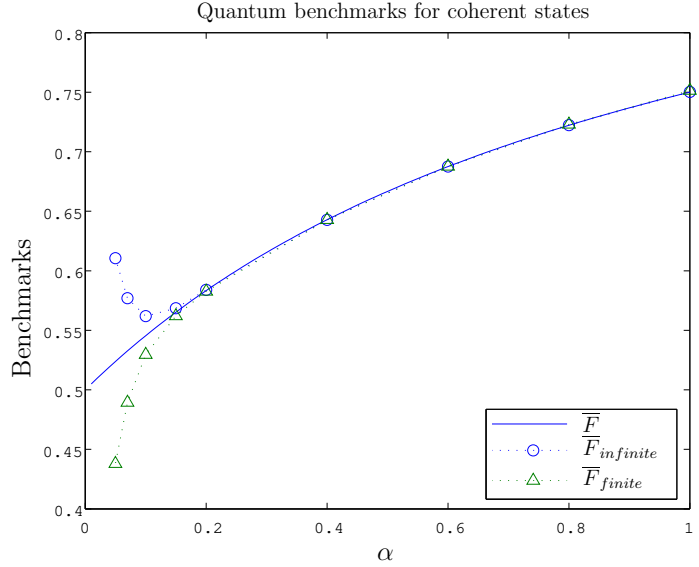,angle=0,width=0.7\linewidth}
\vspace*{-0.4cm}
\end{center}
\caption{A plot of the known optimal classical bound
$\overline{F} = \frac{2\alpha  +1}{2(\alpha +1)}$,
the upper bound $\overline{F}_{infinite}$ derived by our numerical calculation with $c = 35$,
and the result of the finite SDP $\overline{F}_{finite}$ (without corrected truncation error)
for an ensemble of coherent states, no noise $(\lambda=1)$,
and $q(\xi)= \frac{\alpha}{\pi}\exp [-\alpha ||\xi||^2]$.}
\label{graph2}
\end{figure}

Our computation proceeds along the following steps:
\begin{enumerate}
\item Compute matrix elements of $\bra{k_1} {\mathcal N}_\lambda( \rho_{s,\xi})  \ket{l_1}$ and
$ \bra{k_2}  \rho _{s,\xi}
\ket{\l_2}$ for all $k_1$, $k_2$, $l_1$, $l_2$ satisfying $k_1 + k_2 \le c$ and $l_1 + l_2 \le c$; $c$ is a fixed maximum photon number practically bounded by the computer memory.
For the calculation, we used an analytical formula for matrix elements of single-mode Gaussian states as a {\it finite sum} over
 Hermite polynomials (see Eq.(4.10) of \cite{A95}).
\item Evaluate the following integral over $\xi$
\begin{equation*}
\int _{\xi \in \mathbb{R}^2}
  q(\xi) \bra{k_1} \mathcal{N}_{\lambda}( \rho_{s,\xi} )\ket{l_1} \bra{k_2}  \rho _{s,\xi}\ket{l_2} \; d\xi
\end{equation*}
by Quasi-Monte Carlo with the Halton sequence \cite{Quasi Monte Carlo}.
\item From Eq.(\ref{eq: eta block diagonal}), we derive $P_c\eta P_c$.
\item Finally, we solve the SDP leading to $\overline{F}_{infinite}$.
\end{enumerate}
Here, all the above numerical calculations were implemented in Matlab, and the SDP
is solved using the Matlab toolbox SeDuMi version 1.1 \cite{SeDuMi}.
In the above numerical calculation,
we do not use any approximation except for the Quasi-Monte Carlo integral.

\begin{figure}[ttt]
\begin{center}
\epsfig{file=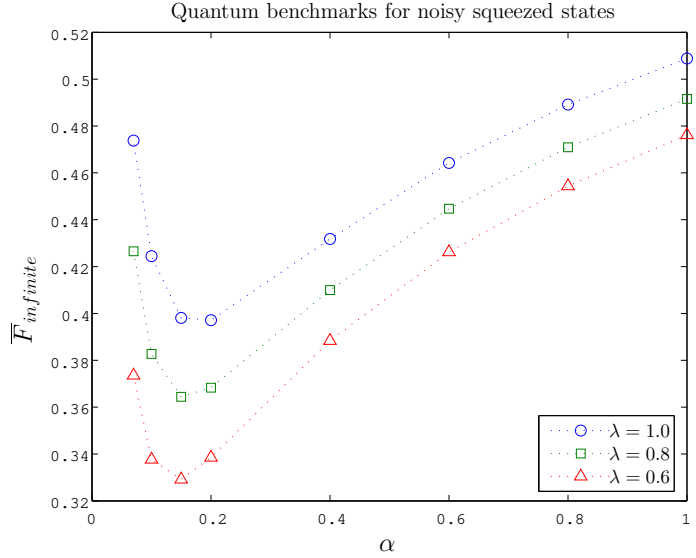,angle=0,width=0.7\linewidth}
\vspace*{-0.4cm}
\end{center}
\caption{Results of numerical calculations of the quantum benchmark $\overline{F}_{infinite}$
for an ensemble of noisy squeezed states displaced according to a
Gaussian probability distribution $q(\xi)= \frac{\alpha}{\pi}\exp [-\alpha ||\xi||^2]$. Levels used in Fock space
 $c = 30$, squeezing $s= 8$, and losses $(1-\lambda) = 0.4, 0.2, 0$.}
\label{graph1}
\end{figure}

As a first application, in order to provide the reader with a convincing test for
our numerical calculations, we show that our method reproduces
the result of the optimal quantum benchmark for displaced coherent states derived in \cite{Hammerer}.
Suppose $\lambda =s=1 $ and the input ensemble is distributed according to $q(\xi)= \frac{\alpha}{\pi}\exp [-\alpha ||\xi||^2]$.
That is we deal with an ensemble of coherent states whose centers are distributed according to a Gaussian distribution with variance $1/(2\alpha)$.
Hence, using the invariance of the ensemble under rotations we obtain
\be
 \eta = \int _{\xi \in \mathbb{R}^2}
   \frac{\alpha}{\pi}\exp [-\alpha ||\xi||^2] \ \rho_{1,\xi}
\otimes \rho _{1,\xi}\;  d\xi \; .
\ee
The quantum benchmark $\overline{F}$ in this case was shown \cite{Hammerer} to be
(note that our definition of the parameter $\alpha$  is different by the factor $2$ from the definition in \cite{Hammerer})
\begin{equation}
\overline{F}  = \frac{2\alpha  +1}{2(\alpha +1)}.
\end{equation}
So we can compare our upper bound $\overline{F}_{infinite}$ derived by SDP with the optimal bound, which is shown in Figure \ref{graph2}
for the maximum photon number $c=35$.
As expected the numerics satisfies $\overline{F}_{infinite}
> \overline{F} >\overline{F}_{finite}$ but we observe in
Fig. \ref{graph2} that as long as $\alpha \ge 0.2$, the
values of both $\overline{F}_{infinite}$ and
$\overline{F}_{finite}$ are essentially indistinguishable
from the optimal bound $\overline{F}$ while for $\alpha \le 0.2$
we find noticeable differences.
In other words, in Eq.(\ref{eq:F _infinite}) the term $\epsilon_{error}$, which originates from the truncation of the dimension,
becomes dominant in this region.
As a result, the minimum value of $\overline{F}_{infinite}$ is around $\alpha = 0.15$, i.e., $\overline{F}_{infinite}$ is not monotonically decreasing with respect to $\alpha$.

From this result, we may expect that our upper bounds are also almost optimal in other situations (different $q,s,\mathcal N$)
as long as $\epsilon_{error}$ is small with respect to $\overline{F_{finite}}$.

Next, we consider the case of an ensemble of noisy squeezed states with a fixed squeezing parameter $s$ and again
a Gaussian prior distribution for $\xi$.
We show the results of our calculation in Figure \ref{graph1}.
Here, we chose $c = 30$, $s = 8$, and plot for $\lambda = 0.6, 0.8, 1$.
In this figure, we can observe that, as expected, $\overline{F}_{infinite}$
decreases with decreasing $\lambda$.
Due to the increasing contribution of $\epsilon_{error}$ for decreasing $\alpha$,
the best (lowest) values of our benchmarks
are achieved, under realistic losses, for $\alpha\simeq 0.15$.
This value, though maybe non-optimal, is suitable for comparisons with
realistic experimental situations.
A simple, preliminary analysis of possible experimental noise conditions indicates that
the values of the benchmark for such parameters could be beaten by
forthcoming experiments aimed at the teleportation or storage of squeezed
states.
Let us mention that single states have already been teleported or stored,
both with light modes \cite{singleteleport1,singleteleport2} and with atomic memories \cite{squeezedmemory1,squeezedmemory2}.
The experimental demonstration of the transmission of a full ensembles of squeezed states
is yet to come, but techniques are ripe for it to be realized in both settings:
our method is ready for the analysis of such developments by direct comparison.

\begin{figure}[ttt]
\begin{center}
\epsfig{file=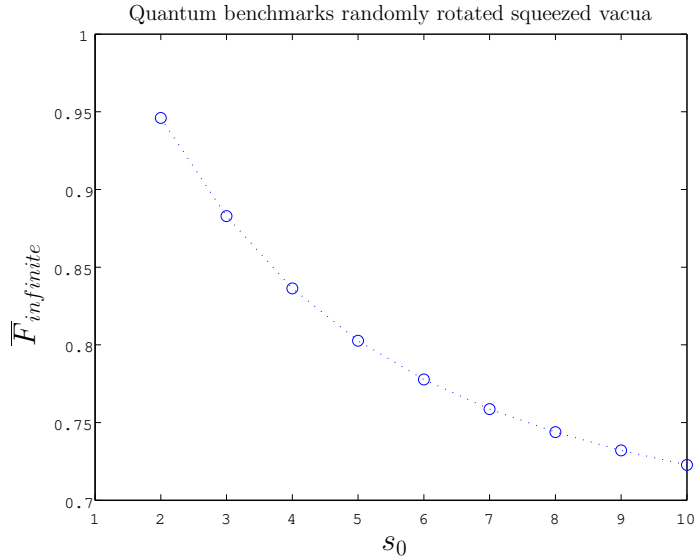,angle=0,width=0.7\linewidth}
\vspace*{-0.4cm}
\end{center}
\caption{The results for the quantum benchmark $\overline{F}_{infinite}$
for an ensemble of randomly rotated squeezed vacuum states without displacement, i.e., $q(\xi)=\delta(||\xi||^2) $,
 $c = 30$ and no noise ($\lambda = 1$).}
\label{graph4}
\end{figure}

Finally, we consider the  simple case of an ensemble of  randomly rotated squeezed vacuum states
without displacement in phase space.
That is, we choose $q(\xi)=\delta(||\xi||^2) $, shown in Figure \ref{graph4}.
Here, we chose $c = 30$  and plot for $2 \le s \le 10$.
Note that for $s=1$ there is only one state in the ensemble, the vacuum state, so that $\overline{F}=1$ in this case.
Fig.~\ref{graph4} clearly shows that, if the displacements in phase space are not randomised,
the classical benchmarks we derived
increase significantly and and beating them in current experiments would be a daunting
challenge:
distributing the displacements of the input ensemble
is thus definitely, at present, a technical necessity
in order to achieve a certifiable quantum performance \cite{miguel}.


\section{Summary and Conclusions}
In this work we have derived upper bounds on the performance
of classical 'measure and prepare' protocols
for the quantum storage and teleportation
of ensembles of squeezed states of continuous variable systems.
These bounds may be employed as
benchmarks to discern truly quantum mechanical performances
in experiments aimed at the realization of such processes.
Motivated by currently available experimental capabilities and
concrete set-ups, we have considered ensembles comprised of states with fixed squeezing,
but allowed for the possibility of random rotations and
translations of those states in phase space.

Fully analytical benchmarks have been obtained
for pure squeezed states as well as mixed states obtained
from the application of additive thermal noise, for the case of completely
unknown rotations and translations in phase space.
In these ideal cases, the benchmarks decrease 
monotonically with increasing degree of squeezing $s$, 
vanishing for diverging $s$: in this sense, 
increasing the squeezing reveals the infinite dimensional 
character of the Hilbert space.
Furthermore, we have presented
a numerical technique that first reduces the
problem to a finite dimensional setting with rigorous error
bounds, and then allows for the numerical solution of the
remaining finite dimensional problem using semi-definite
programming. The so obtained bounds are rigorous and this
approach may be applied to more general settings, even
beyond the Gaussian regime, for variable squeezing and
arbitrarily distributed displacements, with the only proviso that
the distribution over rotations should be still rotationally symmetric in phase space,
to maintain covariance under rotations
(which was essential in deriving rigorous bounds for the truncated problem).
Let us also emphasise that our numerical strategy can be reliably applied to 
finite sets of states as well (rather than to continuous distributions), 
which are what, strictly speaking, is sampled in actual experiments.

Our numerical results strongly indicate that allowing for randomised displacements
is crucial to outperform optimal classical strategies with current quantum technologies
and that, on the other hand,
if such a randomisation is allowed, realistic setups might be able to enter the quantum
regime with squeezed states.
Likewise, randomised phases, other than providing a much more ``appealing'' evidence for
the teleportation or storage of the states (since, in such a case, states with a varying
structure of second moments would be transmitted),
seems to be needed for the benchmarks to be beaten.

Overall, the results presented here will provide a useful resource for assessing
the quantum mechanical character of
experiments aimed at demonstrating quantum storage and
transmission of squeezed states. In a future work, these methods will be
applied to concrete experiments.

\bigskip

\noindent{\em Note added.} We acknowledge that, very recently
(after the completion of this study and
during the writing of the present paper),
another work appeared (Ref.~\cite{calsamiglia})
where benchmarks based on the Uhlmann fidelity for
rotationally covariant ensembles were studied, 
and the reduction of the benchmark estimation to an SDP was independently 
derived.

\ack
\noindent We warmly thank B. Melholt Nielsen and T. Fernholz for 
discussions. This work was supported by the EU Integrated Project 
QAP, the EU STREPs COMPAS, CORNER, HIP and QUANTOP, the EPSRC grant 
EP/C546237/1, the EPSRC QIP-IRC, a Royal Society Wolfson Research 
Merit Award, and the Danish Natural Science Research Council (FNU).

\appendix

\section{Phase space covariant channels}\label{phspcovariant}
In this appendix, we collect useful results about `phase space covariant' channels,
{\em i.e.} channels covariant under the action of the Weyl-Heisenberg group of displacement
operators $\{W_{\xi}, \xi\in{\mathbb R}^{2n}\}$. For a phase space covariant channel one has
$$
T(W_{\xi}^{\dag} \hat{O} W_{\xi}) = W_{\xi}^{\dag} T(\hat{O}) W_{\xi} \, , \quad \forall \,
\hat{O}\in {\B}(L_{2}({\mathbb R})^{\otimes n}) \quad {\rm and} \quad
\forall\,\xi\in\mathbb{R}^{2n} \, .
$$
[${\B}(L_{2}({\mathbb R})^{\otimes n})$ being the set of bounded linear operators on
$n$ copies of the bosonic Hilbert space].
These results were
already used (though not explicitly detailed) in
\cite{OleCloning,OleDiplom}.
Central to this analysis is the characterization of
the class of ``linear bosonic channels'' in terms of their action on
Weyl operators, given in \cite{DV}, which we report here without proof
in the form of the following lemma. Recall that $T^*$ stands for the operation $T$ in the Heisenberg picture.
\begin{Lemma}[Linear bosonic channels]\label{lem:linboschannels}
A map $T^*(W_\xi):=f(\xi)W_{X\xi}$ is a quantum channel if and only if $f$ is
the quantum characteristic function with respect to a modified symplectic
form $\tilde\sigma:=\sigma-X^T\sigma X$. That is,
$f:\mathbb{R}^{2n}\rightarrow\mathbb{C}$ has to be continuous,
$f(0)=1$ and every matrix with entries
\be\label{BKmodified}M_{kl}=
f\big(\xi^{(k)}-\xi^{(l)}\big)\exp{\Big[\frac{i}{2}\xi^{(k)}\cdot\tilde\sigma\xi^{(l)}\Big]}\ee
has to be positive semi-definite for all
$\xi^{(k)},\xi^{(l)}\in\mathbb{R}^{2n}$.
\end{Lemma}
Note that
condition (\ref{BKmodified}) for $\tilde\sigma=\sigma$ ($\tilde\sigma=0$)
is equivalent to $f$ being a quantum
(classical) characteristic function.

The following characterization of phase space covariance ensues:
\begin{Lemma}[Phase space covariant
channels]\label{lem:Wcovariant}
Every phase space covariant channel is uniquely characterized by a
classical characteristic function $f$ and acts in the Heisenberg
picture as \be T^*(W_\xi) =f(\xi) W_\xi.\ee
\end{Lemma}
\begin{Proof}{Consider the action of $T^*$ on a Weyl operator $W_\xi$.
Exploiting the Weyl relations, phase space covariance (with respect to
any $W_\eta$) leads to
\be W_\eta
T^*(W_\xi)=e^{i\xi\cdot\sigma\eta}\; T^*(W_\xi) W_\eta \, .
\ee
It is straightforward to check that this entails
\be
\big[T^*(W_\xi)W_\xi^\dagg,W_\eta\big]=0 \, , \quad \forall\, \xi,\eta\in\R^{2n} \, .
\ee
The irreducibility of the Weyl
system then implies $T^*(W_\xi)W_\xi^\dagg\propto\1$. Denoting
the proportionality constant by $f(\xi)$, we have a map of the form
in Lemma \ref{lem:linboschannels} with $X=\1$. Hence, $f$ has to be
a classical characteristic function. }
\hfill $\square$
\end{Proof}

Let us denote by $\vartheta$ the time reversal (or matrix transposition) operator.
Every entanglement breaking channel, {\em i.e.}, `measure-and-prepare
scheme' (see Ref.~\cite{HolevoEB}), is such that
$ T\circ\vartheta$ is completely
positive (that is, the Choi matrix -- or the Jamiolkowski state -- has positive
partial transpose).
Moreover, time reversal acts very simply in phase space by flipping the sign
of one of the two canonical quadratures (see, {\em e.g.}, \cite{simon00}):
$\vartheta(W_{\xi}) = W_{Z\xi}$, where $Z=\oplus_{j=1}^{n} \sigma_{z}$
($\sigma_{z}$ being the Pauli $z$ matrix).
We will now apply this additional constraint to
phase space covariant channels, in order to achieve
a stronger characterization:
\begin{Lemma}[Time-reversible channels]\label{lem:timerev}
A phase space covariant channel $T$ is such that $T\circ\vartheta$ is
completely positive iff it has the form
\be
T^*(W_\xi)=f\big(\xi/\sqrt{2}\big)W_\xi,
\ee
where $f$ is any
quantum characteristic function, {\em i.e.}, there is a density operator
$\tau\in{\mathfrak C}_1(L_{2}(\R)^{\otimes n})$ such that $f(\xi)=\Tr{\left[\tau W_\xi\right]}$.
\end{Lemma}
\begin{Proof}
{Using Lemma \ref{lem:Wcovariant} we obtain
\be
\vartheta\circ T^*(W_\xi)=f(\xi)\vartheta(W_\xi)=f(\xi)W_{Z\xi} \, ,
\ee
where $Z$ reverses the momenta so that $Z^T\sigma Z=-\sigma$.
Following Lemma \ref{lem:linboschannels} this map is completely
positive if and only if positivity of Eq.~(\ref{BKmodified}) holds for
$\tilde\sigma=2\sigma$. The result follows then by substituting
$\xi\rightarrow\xi/\sqrt{2}$.}
\hfill $\square$
\end{Proof}


\section{Proof of Theorem \ref{theorem isomorphism}}\label{proof of isomorphism}
\begin{Proof}
Suppose $\Psi$ is a entanglement breaking channel and can be written \cite{HolevoEB,EB} as $\Psi(\rho) = \int _{\cal X} \Tr(M(dx)\rho)\sigma(x) $
for all $\rho \in \mathfrak{C}_1(\Hi)$, where  $\sigma(x) \in \mathfrak{C}_1(\Hi)$ are states and $M$ is a POVM, i.e., $M(X)\geq 0$ for all Borel subsets $X$ of a complete separable metric space $\cal X$ for which $M({\cal X})=\1$.

Then, $\Omega(\Psi) = \int _{\cal X} M(dx) \otimes \sigma(x)$  satisfies $Tr_B (\Omega(\Psi)) = \1_A$ and Eq.(\ref{isomorphism}).
 Moreover, suppose there exists a $\Omega'(\Psi)$ satisfying $Tr_B (\Omega'(\Psi)) = \1_A$ and Eq.(\ref{isomorphism}).
Then, since $\Tr (\Omega'(\Psi) \ket{i}\bra{j}\otimes \ket{k}\bra{l})=
\Tr(\ket{k}\bra{l}\Psi (\ket{i}\bra{j})) = \Tr (\Omega(\Psi) \ket{i}\bra{j}\otimes \ket{k}\bra{l})$
for a orthonormal basis $\{ \ket{i} \}_i$ of $\Hi$, we  immediately have $\Omega'=\Omega$.

Conversely, suppose there exists a separable positive bounded operator $\Omega$ on $\Hi \otimes \Hi$ satisfying $Tr_B (\Omega) = \1_A$ and
can be written as $\Omega = \int |\Phi\rangle\langle\Phi|\otimes|\varphi\rangle\langle\varphi|\; \mu(d\Phi d\varphi)$ with  measure $\mu$.
We define a channel $\Psi$ as $\rho\mapsto\int \Tr(M(d\varphi)\rho)|\varphi\rangle\langle\varphi|$, where $M(d\varphi) \stackrel{\rm def}{=}\int_\Phi|\Phi\rangle\langle\Phi|  \mu(d\Phi d\varphi) $.
Then, since $Tr_B (\Omega) =  \1_A$, $ M$ is a POVM; that is, $\Psi$ is an entanglement breaking channel.
Evidently, $\Psi$ satisfies Eq.(\ref{isomorphism}).
Moreover, suppose an entanglement breaking channel $\Psi'$ also satisfies Eq. (\ref{isomorphism}).
Then, since $\Tr(B\Psi' (A)) = \Tr (\Omega(\Psi) A\otimes B) = \Tr(B\Psi (A)) $ for all $B \in \B(\Hi)$,
$\Psi'(A) = \Psi(A) $ for all  $A \in \mathfrak{C}_1(\Hi)$.
\hfill $\square$
\end{Proof}





\begin{thebibliography}{99}
\bibitem{bound_note}{An alternative way of noting that this bound remains the same when
restricting to measure-and-prepare schemes is to realize that (i)
$\tau=|0\rangle\langle 0|$ (a Gaussian!) yields the maximum in
Eq.(\ref{eq:suptauinfrho}) and (ii) for two-mode Gaussian states
(bearing in mind the Choi-matrix of $T$) the PPT-criterion is
necessary and sufficient for separability.}
\end{thebibliography}
\end{document}